# Robust A-type order and spin-flop transition on the surface of the antiferromagnetic topological insulator MnBi$_2$Te$_4$

Paul M. Sass[1], Jinwoong Kim[1], David Vanderbilt[1], Jiaqiang Yan[2], and Weida Wu[1]*.

[1]*Department of Physics and Astronomy, Rutgers University, Piscataway, NJ 08854, USA.*

[2]*Materials Science and Technology Division, Oak Ridge National Laboratory, Oak Ridge, Tennessee 37831, USA.*

*Correspondence to: wdwu@physics.rutgers.edu (WW).

**Abstract**

Here we present microscopic evidence of the persistence of uniaxial A-type antiferromagnetic order to the surface layers of MnBi$_2$Te$_4$ single crystals using magnetic force microscopy. Our results reveal termination-dependent magnetic contrast across both surface step edges and domain walls, which can be screened by thin layers of soft magnetism. The robust surface A-type order is further corroborated by the observation of termination-dependent surface spin-flop transitions, which have been theoretically proposed decades ago. Our results not only provide key ingredients for understanding the electronic properties of the antiferromagnetic topological insulator MnBi$_2$Te$_4$, but also open a new paradigm for exploring intrinsic surface metamagnetic transitions in natural antiferromagnets.

Recent progress in topological quantum materials suggest that antiferromagnets may host interesting topological states [1]. For example, it has been proposed that an axion insulator state with topological magnetoelectric response could be realized in an antiferromagentic topological insulator (TI) phase [2,3], where the $Z_2$ topological states are protected by a combination of time-reversal symmetry and primitive-lattice translation. The antiferromagnetic TI state adiabatically connects to a stack of quantum Hall insulators with alternating Chern numbers [4], thus providing a promising route to realizing the quantum anomalous Hall (QAH) effect in a stoichiometric material. The prior observation of the QAH effect in magnetically doped TI thin films is limited to extremely low temperature because of the inherent disorder [5–9], though the disorder effects can be partially alleviated by material engineering [10–12]. The $MnBi_2Te_4$ (MBT) family was predicted and confirmed to be an antiferromagnetic TI that may host QAH and axion-insulator states in thin films with odd and even numbers of septuple layers (SLs) respectively [13–17]. Recent transport measurements on exfoliated thin flakes provide compelling evidence for these predictions [18,19], suggesting gapped topological surface states. On the other hand, recent high-resolution angle-resolved photoemission spectroscopy (ARPES) studies reveal gapless (or small-gap) surface states below the antiferromagnetic ordering temperature, suggesting a surface relaxation of the A-type order and/or the formation of nanometer-sized magnetic domains [20–23]. The antiferromagnetic domain structure of $MnBi_2Te_4$ was revealed by imaging of domain walls using magnetic force microscopy (MFM) [24]. The observed domain size is on the order of 10 µm, excluding the speculated nanometer-size domain scenario [22].

Thus, it is crucial to reveal the nature of surface magnetism of $MnBi_2Te_4$ in order to resolve the dichotomy between the observations of QAH transport and gapless topological surface states [18–23]. The magnetic imaging of A-type domain structures in $MnBi_2Te_4$ also enable explorations of the long-sought surface spin-flop (SSF) transition in a natural antiferromagnet [25–31]. In this letter, we report the observation of alternating termination-dependent magnetic signals on the surface of $MnBi_2Te_4$ single crystals using cryogenic MFM, which provides direct evidence of the persistence of uniaxial A-type antiferromagnetic order all the way to the surface. Combined with the recent ARPES observations of gapless surface states, our results suggest a possible scenario of a tiny magnetic mass gap due to weak coupling between the topological electronic states and the magnetic order. The robust A-type order is further corroborated by the observation of two SSF transitions on domains with opposite terminations revealed by the magnetic field dependence of the domain contrast. Although they have been theoretically studied for decades [25,28,29], SSF transitions have only been observed in synthetic antiferromagnets, not in natural ones [26,27,30,31]. Our results not only shed new light on the realization of topological states in antiferromagnets, but also open up exciting explorations of surface metamagnetic transitions in functional antiferromagnets.

For an A-type antiferromagnet with ordered moments along the *c*-axis, there are only two possible domain states, up-down-up-down ($\uparrow\downarrow\uparrow\downarrow$) and down-up-down-up ($\downarrow\uparrow\downarrow\uparrow$). They are related to each other by either time reversal symmetry or a primitive lattice translation, so they are antiphase domains and the antiferromagnetic domains walls separating them are antiphase boundaries. Therefore, there would not be any vertex point connecting three or more domain walls. These expectations are confirmed by our recent cryogenic magnetic force microscopy (MFM) studies in high magnetic fields [24]. The typical domain size is ~10 µm, so the tiny contribution of chiral edge states at domain walls is insufficient to explain the gapless topological surface states [22]. However, it is unclear whether the A-type order persists up to the surface layer, because MFM contrast could come from sub-surface stray fields that penetrate the surface non-magnetic layer [32]. It has been speculated that the observed gapless surface states might be explained by surface relaxation or reorientation of the A-type order [20–22]. To address these issues, we carried out MFM studies on as-grown surface of $MnBi_2Te_4$ single crystals with multiple SL steps and thin layers of surface impurity phase. Prior studies suggest that the as-grown surface of $MnBi_2Te_4$ is decorated with small amounts of impurity-phase $Bi_{2-x}Mn_xTe_3$, which is a soft ferromagnet with a small coercive field (<0.04 T) [17,33]. These magnetically soft thin layers provide an excellent opportunity to probe the screening effects of the speculated relaxed surface magnetic order with enhanced magnetic susceptibility [21].

Platelike single crystals of $MnBi_2Te_4$ were grown out of a Bi-Te flux and have been well characterized by measuring the magnetic and transport properties [17]. The MFM experiments were carried out in a homemade cryogenic magnetic force microscope using commercial piezoresistive cantilevers. MFM tips were prepared by depositing nominally 150 nm Co film onto bare tips using e-beam evaporation. MFM images were taken in a constant height mode with the scanning plane nominally ~100 nm (except specified) above the sample surface [24]. The numerical simulations were performed with the revised Mills model. The reduced surface magnetization causes a pinning of the spin-flop state at the surface [34].

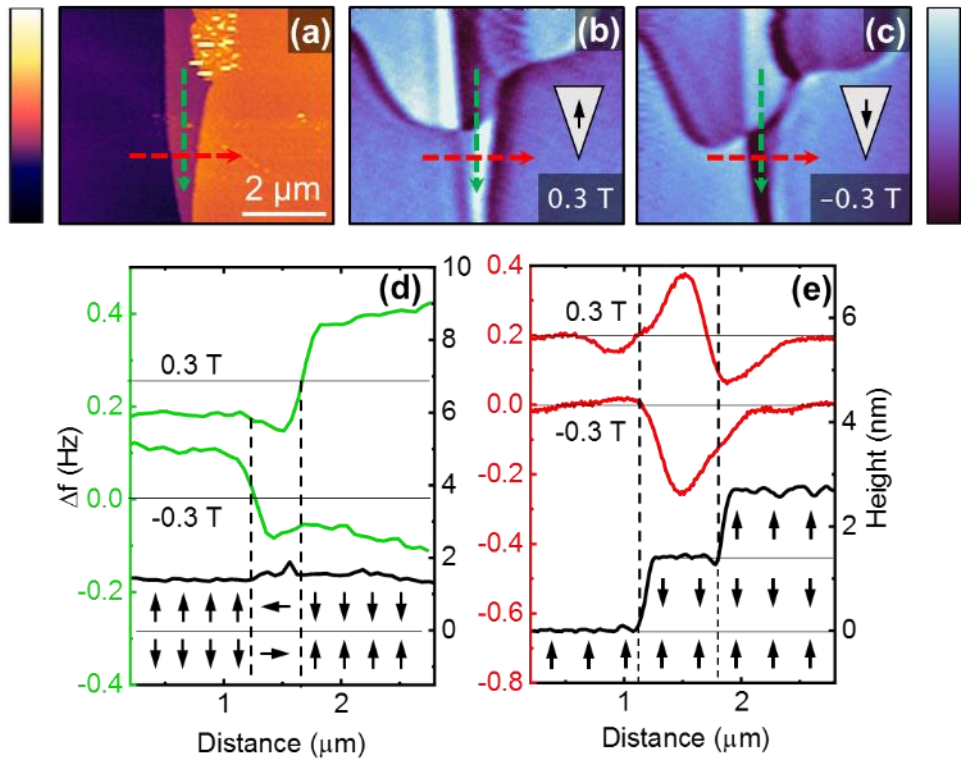

FIG. 1 (a), Topographic image (5 K) of one and two septuple layer (SL) steps on an as-grown $MnBi_2Te_4$ single crystal. (b,c) MFM images taken at 0.3 and −0.3 T, respectively, after field cooling at 0.6 T, at the same location as in (a). The applied magnetic field is perpendicular to the sample surface. A curvilinear domain wall cuts through the SL step. The domain and SL step contrast was reversed when the tip moment was flipped (dark is attractive and bright is repulsive). (d,e), Line profiles of the topography (black) and MFM (green and red) data. The frequency shift in (d) was measured across the domain wall over flat topography, while in (e) it was taken across the SLs. The color scale for the topographic (MFM) image(s) is 6 nm (0.3 Hz).

Figure 1(a) shows a typical surface morphology of MnBi$_2$Te$_4$ as-grown surface. There are two step edges in this location, and the observed step height (~1.3 nm) agrees with that of a single SL. Figs. 1(b) and 1(c) show the MFM images taken at the same location. Note that one antiferromagnetic domain wall cuts across the SL steps. Clearly, the magnetic contrast reverses over the domain wall on one terrace (green arrow) and across SLs of one single domain (red arrow) as shown in Fig. 1(b) and illustrated by line profiles in Fig. 1(d) and 1(e). Here, bright contrast indicates a repulsive interaction, i.e., surface magnetization antiparallel to the MFM tip moment, which is fixed by a small out-of-plane magnetic field [32]. The domain contrast reverses over the domain wall, which is consistent with opposite surface magnetization states of different antiphase domains (Fig. 1(d)) or SL steps (Fig. 1(e)). There is a slight dip at the domain wall due to its higher susceptibility [24]. The slight asymmetry in the line profiles in Fig. 1(e) is due to the difference between forward and backward scanning [34]. The magnetic contrast originates from imperfect cancellation of magnetic stray field from the alternating ferromagnetic layers [35,36]. To confirm this, we reverse MFM tip moment using a negative magnetic field (−0.3 T). The magnetic contrast indeed reverses as shown in Fig. 1(c), which unambiguously demonstrates that the alternating MFM signal is from the alternating surface magnetization. Note that there is a small island of impurity phase (Bi$_{2-x}$Mn$_x$Te$_3$) with a rougher surface sitting on the upper SL step edge (Fig. 1(a)). It appears to screen the antiferromagnetic domain contrast, as shown in Fig. 1(b) and 1(c). To understand the screening effect of the impurity phase, we increase the scan size to sample more impurity phases.

Figure 2(a) shows the topography of a large area with six SL steps in the field of view (~18×13 µm$^2$). Most steps are paired to form curvy narrow terraces decorated with many plate-like impurity islands with partial hexagon shapes. The height of these island (~3 nm) agrees with that of three quintuple layers (QLs) of Bi$_2$Te$_3$, which is slightly larger than that of two SLs (~2.7 nm) as shown in Fig. 2(i) [34]. Fig. 2(b) shows the MFM image (measured at 1 T) at this location after 0.425 T field cooling. There are two bubble-like antiferromagnetic domains with curvilinear domain walls. Alternating magnetic contrast was observed on uncovered SL terraces across step edges or antiferromagnetic domain walls. However, this contrast is suppressed if the surface is covered by the impurity phases, suggesting a very effective screening of the magnetic stray field [34]. To illustrate the details, zoom-in images of a few selected areas (boxes labelled 1, 2 and 3 in Fig. 2(a) and 2(b)) are shown in Fig. 2(c-h). Arrows (dashed lines) marked the exposed (covered) narrow terraces in these images [34]. As shown in box 3, the domain contrast can even be "blocked" by a fractional QL of the impurity phase, and clear domain contrast is visible in the holes of the impurity phase. Thus, we can conclude that the magnetic impurity phase (Bi$_{2-x}$Mn$_x$Te$_3$) effectively screens all the stray fields from the underlying MnBi$_2$Te$_4$ surface. Similar results are observed at higher temperature (below $T_N$). In contrast, antiferromagnetic domain wall contrast is not affected by the impurity phase as shown in the white dotted box in Fig. 2(b), because domain walls extend into the bulk. Because the alternating domain and terrace contrast can be easily screened by such a thin layer (0.3-3 nm) of soft magnet (Bi$_{2-x}$Mn$_x$Te$_3$), the uniaxial A-type spin order must persist to the top surface layer of MnBi$_2$Te$_4$. Otherwise, the termination-dependent magnetic contrast would be screened by any relaxation of surface magnetism with substantial magnetic susceptibility, such as paramagnetism, non-A-type spin order, or in-plane A-type order proposed in prior reports [20–23,37]. Therefore, we can conclude that our MFM observation excludes some of the proposed surface relaxation models, and that the contradictory reports of gapless surface states and a quantized anomalous Hall effect remain unresolved.

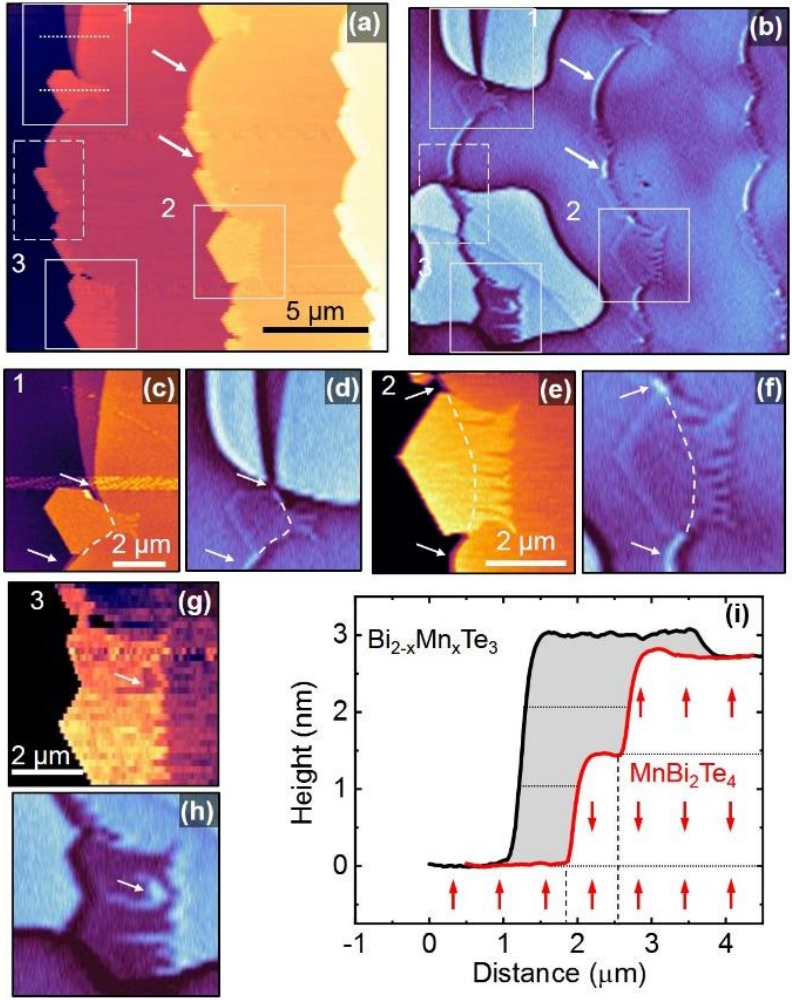

FIG. 2 (a,b) Topographic and MFM images of MnBi$_2$Te$_4$ surface covering measured in 1 T at 5 K after 0.425 T field cooling. Magnetic contrast of domains and terraces is visible. (c-h) Zoom-ins of topographic and MFM images outlined by solid white boxes in (a,b). White arrows (dashed lines) mark the exposed (covered) single SL steps. The bright domain contrast in region covered by the impurity phase is suppressed, as shown by white arrow in (h). Domain wall contrast is not suppressed by the impurity phase, as shown in the dotted box in (b). (i) Topographic line profiles (white dotted lines in (a)) of SLs and impurity phase QLs with schematic of spin configuration. The gray area illustrates a soft magnetic phase that screens the stray fields of the SL edges underneath. The color scales for the topographic and MFM images are 7, 6, 3 and 3 nm (0.2 Hz), respectively.

The observation of robust A-type order on the MnBi$_2$Te$_4$ surface also provides a rare opportunity to explore the interesting SFF transition (or inhomogeneous spin-flop), which was first proposed by Mills decades ago using an effective one-dimensional spin-chain model with AFM nearest-neighbor exchange coupling [25,29]. However, later studies suggested an intriguing scenario of inhomogeneous spin-flop state due to finite size effect [28,30,38]. The SSF transition was observed in synthetic AFMs, which are superlattices of antiferromagnetically coupled ferromagnetic layers [26,27], but not in natural AFMs [28,31]. Because of the existence of domains in natural AFMs, the exploration of SSF phenomena requires a surface-sensitive magnetic imaging probe with sufficient spatial resolution in high magnetic field. These challenges were overcome by our cryogenic MFM.

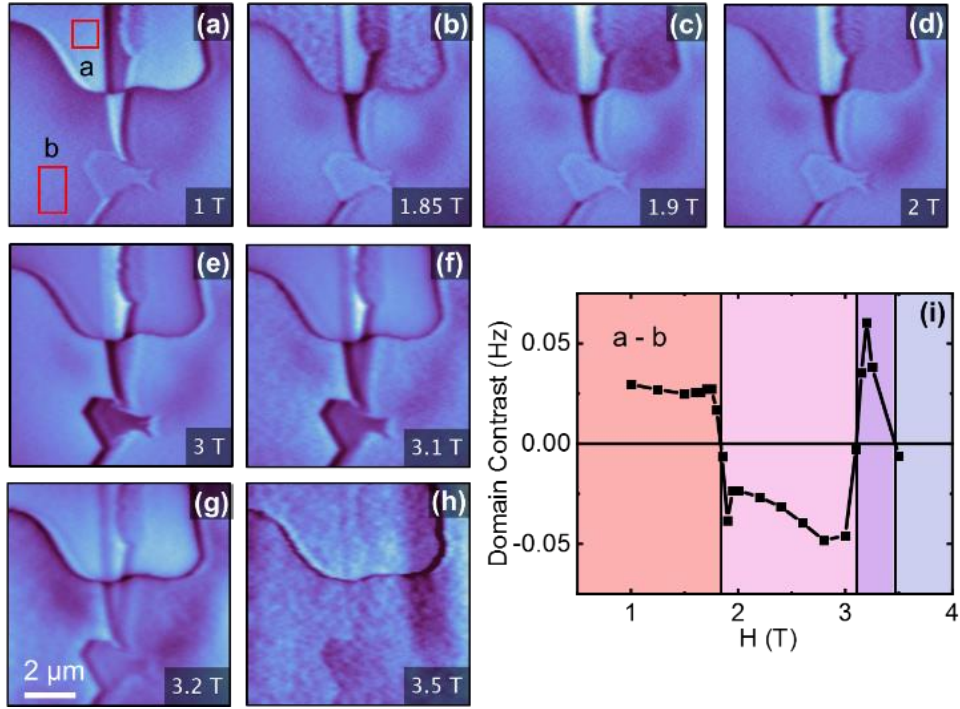

FIG. 3 (a-h) MFM images taken at 5 K with increasing field labeled in lower right corners. (i) Domain contrast between red squares, labeled *a* and *b* in A versus applied field. Below 1.75 T, the domain contrast is constant. As the applied field is further increased, *a* domains starts to appear rougher and darker near 1.85 T, then the domain contrast quickly reverses above 1.85 T. Similar behavior was observed on *b* domains around 3.1 T. Above 3.5 T, the system enters the canted AFM phase and the domain contrast disappears. The color scale for MFM images is 0.3 (a-d) and 0.8 (e-h) Hz.

Figs. 3(a-h) show selected MFM images measured in various magnetic fields from 1.0 to 3.5 T [34]. Clearly, the termination-dependent contrast shows non-monotonic magnetic field dependence. As discussed in connection with Fig. 1, in low magnetic field a bright contrast indicates surface termination with antiparallel magnetization denoted as *a*, while dark contrast indicates surface termination with parallel magnetization denoted as *b* in Fig. 3(a). This domain contrast persists in finite magnetic field up to ~1.85 T, then fine features start to emerge in termination *a* during the domain contrast reversal, while the termination *b* remains featureless. Thus, it is the termination *a* (antiparallel magnetization) that undergoes SSF transition at $H_{SSF}^1$~1.85 T. Similar behavior was observed at ~3.1 T except the roles of *a* and *b* are switched. Thus, it is the termination *b* (parallel magnetization) that undergoes SSF transition at $H_{SSF}^2$~3.1 T. Finally, the domain contrast disappears around the bulk spin-flop (BSF) transition ($H_{BSF}$~3.5$T$). The detailed field dependence of domain contrast is plotted in Fig. 3(i), where the domain contrast is defined as the difference of the average MFM signals in the two regions (domain *a* and *b*) marked by red boxes in Fig. 3(a). This effect is also observed in negative applied field and is reproducible in other sample locations after thermal cycling and on a cleaved crystal of MnBi$_2$Te$_4$ [34]. No hysteresis was found between up-sweep and down-sweep of the magnetic field.

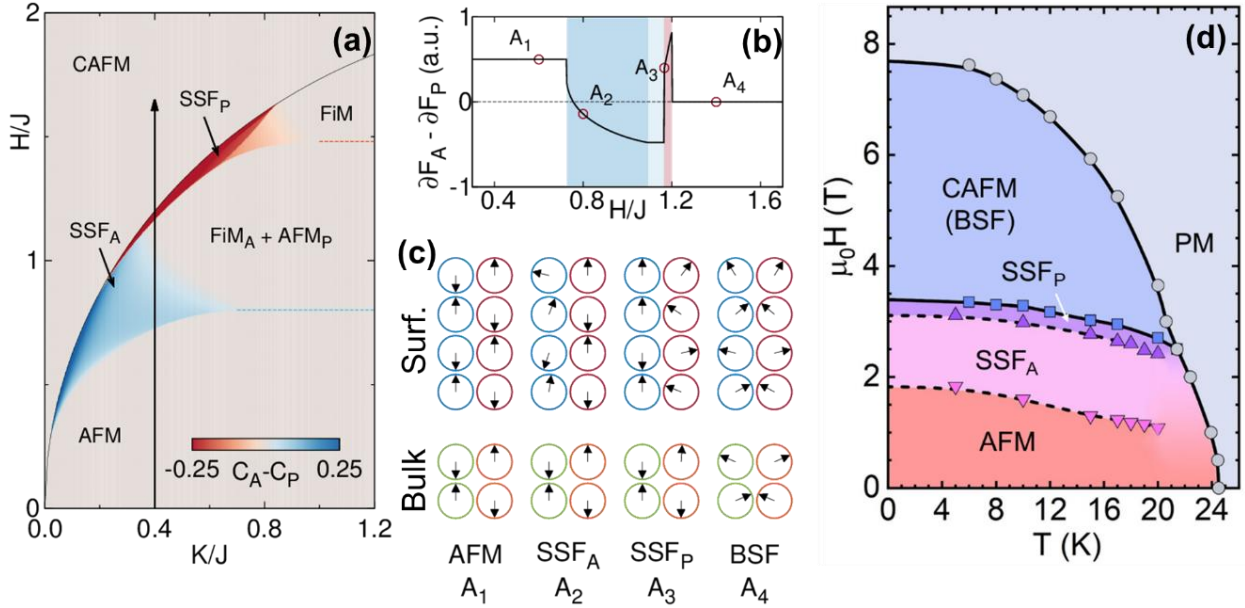

FIG. 4 (a) Theoretical phase diagram of the spin-flop state in the revised Mills model. Blue and red colored regimes illustrate SSF states for antiparallel and parallel surfaces, respectively. Color code denotes the difference of net spin canting between the two types of surfaces [34]. Black solid line is a phase boundary of the bulk spin-flop state; dashed line is a boundary between AFM and SSF phases for antiparallel (blue) and parallel (red) surfaces. (b) Simulated field dependence of magnetic force gradient differences between antiparallel and parallel surfaces. (c) Schematic illustration of the spin-flop process for surface (upper 4 rows) and bulk (lower) domains. Left blue (right red) represents antiparallel (parallel) surface spins, whereas, left green (right yellow) represents antiparallel (parallel) bulk spins. (d) $H$-$T$ phase diagram showing A-type AFM phase (red), $SSF_A$ and $SSF_P$ spin-flop phase (pink and light purple), bulk CAFM phase (dark purple), and forced ferromagnetic or paramagnetic (PM) phase (light blue).

The first SSF transition $\left(H_{SSF}^1 \approx 0.5 H_{BSF}\right)$ agrees well with prior observation in synthetic antiferromagnets [27], and is in reasonable agreement with that of the Mills model $\left(H_{SSF}^{th} \approx 0.7 H_{BSF}\right)$ [29,38]. However, the second SSF transition $\left(H_{SSF}^2 \approx 0.9 H_{BSF}\right)$ of the surface with parallel magnetization is unexpected in prior studies [26,28,38], indicating surface relaxation of the A-type AFM order. To confirm this, we studied the revised Mills model with additional surface relaxation effects such as reduced magnetization, exchange coupling, and/or anisotropy energy [28,34].

In the original Mills model, the antiparallel surface nucleates a horizontal domain wall with a spin-flop state that migrates into the bulk, forming an inhomogeneous state that precedes the bulk spin-flop transition. [28,29,38] If the migration indeed occurs, the antiparallel surface would sequentially turn into a parallel surface, resulting in an identical magnetization state on the two domains, *i.e.*, no domain contrast above the SFF transition. Such behavior is inconsistent with our experimental observation of domain contrast reversal. Our simulation reveals that the horizontal domain wall with spin-flop state can be pinned to surface layers if the magnetization of surface layer is reduced >10% [34]. Indeed, the revised Mills model with surface relaxation effect can reproduce the two successive SSF transitions in a reasonably wide parameter space.

Fig. 4(a) shows a phase diagram of the simulation using typical parameters exhibiting the emergent sequential SSF transitions on antiparallel (blue) and parallel (red) surfaces, respectively. In addition, the reduction of surface exchange coupling could explain the suppression of the SSF transition. The simulated MFM contrast (force gradient difference) as a function of magnetic field is shown in Fig. 4(b), qualitatively agreeing with the experimental observation shown in Fig. 3(i) [34]. The successive SSF and BSF transitions are summarized schematically in Fig. 4(c). The antiparallel surface layer (blue) undergoes a SSF transition $H_{SSF}^1$ where the MFM contrast reverses. The domain contrast increases even further in this region, likely due to an increasing canted moment of the spin-flop state. At the next critical field $H_{SSF}^2$, the parallel surface (red) undergoes SSF transition, resulting in another reversal of the MFM contrast. Finally, the MFM domain contrast disappears above the BSF transition because both domains have the same canted moments.

To explore the impact of thermal fluctuations, we performed MFM studies at higher temperatures below $T_N$ to extract the $T$ dependence of the SSF transitions ($H_{SSF}^1$ and $H_{SSF}^2$) [34]. As shown in Fig. 4(d), the temperature dependence of both SSF transitions follow that of the BSF ($H_{BSF}$), which gradually reduces with increasing temperature until the bicritical point (~21 K, ~2.5 T), indicating the relative energetics of the SSF transitions do not vary much with temperature. Above 21 K, the antiferromagnetic domains become unstable in finite magnetic field because of enhanced thermal fluctuations, making it difficult to determine the SSF transitions in this temperature window.

In summary, our MFM results provide microscopic evidence of robust uniaxial A-type order that persists to the top surface layers in the antiferromagnetic topological insulator MnBi$_2$Te$_4$. Thus, our results strongly constrain the possible mechanisms of the observed gapless topological surface states. Furthermore, we observed, for the first time, the long-sought SSF transition in natural antiferromagnets. More interestingly, we discovered an additional surface SSF on the parallel magnetization surface, which indicates surface relaxation of the A-type order. The MFM observation of the SSF transition not only opens a new paradigm for visualizing surface metamagnetic transitions in antiferromagnetic spintronic devices, but also provides new insights into the realization of the quantum anomalous Hall or axion-insulator states in topological anitferromagnets [18,19].

**Acknowledgement**

The MFM studies at Rutgers is supported by the Office of Basic Energy Sciences, Division of Materials Sciences and Engineering, US Department of Energy under Award numbers DESC0018153. The simulation efforts is supported by ONR Grants N00014-16-1-2951. Work at ORNL was supported by the US Department of Energy, Office of Science, Basic Energy Sciences, Materials Sciences and Engineering Division.